\documentclass[aps,prb,twocolumn,superscriptaddress]{revtex4}

\usepackage{color}
\usepackage{bm}
\usepackage{amsmath}
\usepackage{amssymb}
\usepackage{amsthm}
\usepackage{graphicx}
\usepackage{epsfig}
\usepackage{natbib}
\usepackage{verbatim}

\begin{document}
\title{Intersubband polaritons with spin-orbit interaction}
\author{O. Kyriienko}
\affiliation{Science Institute, University of Iceland, Dunhagi 3,
IS-107, Reykjavik, Iceland}
\affiliation{Division of Physics and Applied Physics, Nanyang Technological University 637371, Singapore}
\author{I. A. Shelykh}
\affiliation{Science Institute, University of Iceland, Dunhagi 3,
IS-107, Reykjavik, Iceland}
\affiliation{Division of Physics and Applied Physics, Nanyang Technological University 637371, Singapore}
\date{\today}

\begin{abstract}
We investigate intersubband polaritons formed in the asymmetric quantum well (AQW) embedded into the semiconductor microcavity and study the effects of spin-orbit interaction (SOI) acting on intersubband excitations. The spin-orbit interaction of Rashba and Dresselhaus type remove the spin degeneracy of electrons with finite value of in-plane momentum and allow four types of intersubband excitations. While optical spin-flip transitions are suppressed, the spectrum of elementary excitations shows the appearance of upper, lower and middle polaritonic branches based on spin-conserving transitions. The accounting of finite photon momentum leads to non-zero average spin projection of electronic ensemble in the first excited subband under cw excitation for both isotropic (Rashba) and anisotropic (Rashba and Dresselhaus) SOI. We predict the possibility of spin current generation in the considered systems with long coherence length.
\end{abstract}
\maketitle

\section{Introduction}

Intersubbband transition in semiconductor quantum well (QW) plays significant role in the modern optoelectronics due to numerous possible applications in optical devices operating in the infra-red and terahertz frequency domains.\cite{Dupont2003,Walther,Cathabard,NatureComm} The dependence of energy distance between subbands on QW width allows to
adjust the frequency of photon emitter or detector in relatively easy way comparing to the usual interband transition. Moreover, the implementation of multiple QW samples gives the possibility to create devices with high efficiency, in particular quantum cascade lasers.\cite{Faist,Colombelli}

An important characteristic of intersubband transition as compare to interband transitions is the peculiar optical selection rules which only allow absorption of TM polarized electromagnetic mode because dipole element of transition for TE polarized mode is zero.\cite{Liu} Furthermore, it is possible to improve the efficiency of light interaction with absorbing media by placing it into the semiconductor microcavity. This allows to achieve the strong coupling regime when in case of intersubband transition cavity photons are constantly absorbed and emitted and mixed light-matter modes are formed.\cite{Dini,Geiser,Kyriienko}

Up to now, a consideration of the spin properties for intersubband polaritons was never performed. On the other hand, the spin electronics or \textit{spintronics} is one of the most fastly developing areas of mesoscopic physics. The main issue of
spintronics is spin transfer in the system and generation of spin currents. Being widely studied nowadays, it proposes the devices which operate on other principles comparing to usual electronics,\cite{Wolf,Zutic} for instance -- the spin field effect transistor.\cite{Datta}

One should note, that contrary to the case of the intersubband polaritons, the spin properties of interband cavity polaritons were widely studied.\cite{ShelykhReview} Moreover, by analogy with spintronics, its optical counterpart, \textit{spinoptronics}, became recently an area of the intensive studies.\cite{Spinoptronics} In this domain the role of Rashba SOI is played by so-called TE-TM splitting\cite{BerryPhase} and optical analogs of various spintronic components were theoretically proposed and experimentally realized.\cite{PolaritonDevices}

One of the basic concepts in spintronics is the spin-orbit interaction (SOI)\cite{Winkler,Silsbee} which appears in semiconductors due to intrinsic bulk inversion asymmetry (BIA) or structure inversion asymmetry (SIA).

The part of SOI appearing in the systems with structure inversion asymmetry, \textit{e.g.} asymmetric quantum wells (AQW), is known as Rashba term. It can be represented by introduction of the in-plane effective magnetic field causing the precession of the electron spin.\cite{Rashba} The corresponding Hamiltonian reads
\begin{equation}
H_{SIA}=\alpha(\sigma_{x}k_{y}-\sigma_{y}k_{x})=\frac{\hbar}{2}(\mathbf{\Omega}_{SIA}\cdot
\mathbf{\sigma}), \label{HRashba}
\end{equation}
where $\alpha$ is a Rashba SOI constant and $\mathbf{\Omega}_{SIA}=2\alpha \hbar^{-1}(k_y;-k_x)$ denotes the
effective magnetic field (measured in frequency units). The diagonalization of the electron Hamiltonian accounting for Rashba SOI gives modifies dispersions where two different spin components are split in energy for non-zero values of electron in-plane momenta. This removes the spin degeneracy in the system and makes possible effective spin control due to the possibility to tune the Rashba coupling parameter $\alpha$ by external gate voltage applied perpendicular to structures
interface.\cite{AlphaTune,AlphaTune1,AlphaTune2,AlphaTune3}

Spin-orbit interaction arising from the bulk inversion asymmetry (BIA) is known as the Dresselhaus term. Similarly to the Rashba SOI, it leads to the appearance of linear in $k$ effective magnetic field oriented in plane of the QW, but has different symmetry. The corresponding Hamiltonian can be written as\cite{Dresselhaus}
\begin{equation}
H_{BIA}=\beta(\sigma_{x}k_{x}-\sigma_{y}k_{y})=\frac{\hbar}{2}(\mathbf{\Omega}_{BIA}\cdot
\mathbf{\sigma}), \label{HDresselhaus}
\end{equation}
where $\beta$ is Dresselhaus constant for material, $\mathbf{\Omega}_{BIA}=2\beta \hbar^{-1}(k_x;-k_y)$. In realistic QWs
usually both types of spin splitting are present. This leads to strongly anisotropic pattern of the effective magnetic field acting on electron spin in the reciprocal space.

In this article we study the effect of Rashba and Dresselhaus SOI on the intersubband optical transitions and formation of intersubband polaritonic states. Electrons in energy subbands of asymmetric QW are subjected to SIA and BIA spin-orbit interaction and both fundamental and upper subbands are spin split for $k\neq0$. This opens four different optical transitions and allows to form five different polaritonics states. In the article we show that optical transitions with spin-flip are suppressed in semiconductor microcavity and spin-conserving excitations interacting with cavity photon give birth to three strongly coupled polaritonic modes with peculiar spin polarization. Tuning the pump conditions one can
generate the spin currents with long coherence length.

The article is organized as follows. In Sec. \ref{subsec:model} we present a Hamiltonian for intersubband excitation subjected to Rashba and Dresselhaus spin-orbit interaction. In Sec. \ref{subsec:polarization} the possibility of optical spin orientation due to non-zero photon momentum is discussed. In Sec. \ref{subsec:current} we discuss spin currents generation by linearly polarized light for intersubband transition. In Sec. \ref{sec:polaritons} we introduce the strong light-matter coupling in the system and show the spectrum of excitation with corresponding spin polarization of polaritonic modes. Finally, Sec. \ref{sec:conclusions} summarizes the results of the article.

\section{Photoabsorption of individual QW with SOI, spin polarization and spin currents}

\subsection{Model Hamiltonian}
\label{subsec:model}

\subsubsection{Rashba SOI}

\begin{figure}
\includegraphics[width=0.7\linewidth]{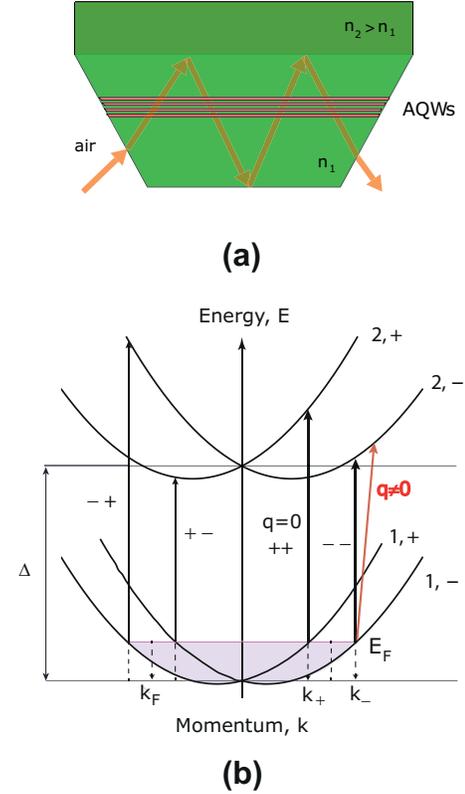}
\caption{(Color online) (a) Sketch of the  geometry of the system. The set of asymmetric QWs is bounded into cavity formed due to the effect of total internal reflection (waveguide geometry). (b) The schematic representation of four possible transitions between upper and lower subbands splitted by SOI. $\Delta$ denotes bare transition energy, $k_{+}$ and $k_{-}$ show the radii of Fermi surface for different SOI subbands.}
\label{Fig1}
\end{figure}

We consider a system of asymmetric GaAs/AlGaAs quantum wells embedded into the microcavity, where TM polarized cavity photons are confined (Fig. \ref{Fig1}(a)). TE polarized mode can be excluded from consideration, as it is not coupled to intersubband transition in the dipole approximation. As compared to the case of symmetric rectangular quantum well, the asymmetry introduces Rashba spin-orbit interaction which induces the spin flips for the electrons moving with finite value of the momentum $k$.

The generic Hamiltonian for the considered system in secondary quantization representation can be written as
\begin{align}\label{H}
&H=\sum_{k,j,\sigma}E_{k,j}a^{\dagger}_{k,j,\sigma}a_{k,j,\sigma}+
\sum_{k,j}[\alpha^{R}_{j}(k_{y}+ik_{x})a^{\dagger}_{k,j,\uparrow}a_{k,j,\downarrow}+ \\ \notag &+
H.c.]+\sum_{q}E^{ph}_{q}b^{\dagger}_{q}b_{q}+\sum_{k,q,\sigma}(g_{q}a^{\dagger}_{k,1,\sigma}a_{k+q,2,\sigma}b^{\dagger}_{q}+H.c.),
\end{align}
where we have chosen the axis perpendicular to the interface of the QW as spin quantization axis $z$. $a^{\dagger}_{k,j,\sigma}$, $a_{k,j,\sigma}$ are creation and annihilation operators for electron with wave vector $\mathbf{k}$ and spin $\sigma$ in lower ($j=1$) or upper ($j=2$) subband, $g_{q}$ is the electron-photon interaction constant which originates from dipole matrix element of intersubband transition and can be calculated as\cite{Kyriienko}
\begin{equation}
g_q=\sqrt{\frac{\Delta\cdot d_{10}^{2}}{\hbar^{2} \epsilon \epsilon_{0} L_{cav}A\omega_{0}(q)}\frac{q^{2}}{(\pi/L_{cav})^{2}+q^{2}}},
\label{g}
\end{equation}
where $L_{cav}$ is cavity length, $\Delta$ is separation energy between levels, $\omega_{0}(q)$ denotes cavity mode dispersion, $\epsilon_{0}$ and $\epsilon$ are vacuum permittivity and relative material dielectric constant, respectively, $d_{10}$ stands for the dipole matrix element of the transition and $A$ is an area of the sample.

The first term in Hamiltonian (\ref{H}) describes free particles and the second term corresponds to the Rashba spin-orbit interaction, where $\alpha_{j}$ are Rashba coefficients for different subbands, $j=1,2$. Note, that in general $\alpha_1\neq\alpha_2$.\cite{AlphaTune3} This term contributes to the mixing of $\uparrow$ and $\downarrow$ states. The third term is free cavity photons energy and fourth one describes the interaction between photons and intersubband transition which conserves electrons spin.

It is convenient to diagonalize the electronic part of the Hamiltonian (\ref{H}) introducing the new operators of spin states $\widetilde{\sigma}=+,-$ oriented perpendicular to the direction of the momentum $\textbf{k}$
\begin{equation}
c_{k,\pm}=\frac{1}{\sqrt{2}}\left[\pm ie^{-i\theta_{k}}a_{k,\uparrow}+a_{k,\downarrow}\right],
\label{new_operators_R}
\end{equation}
where $\theta_{k}=\arctan(k_{y}/k_{x})$ is the angle between radial vector $\mathbf{k}$ and $x$ axis. The Hamiltonian of the light-matter coupling written in new operators reads

\begin{align}
&\widetilde{H}=\sum_{k,j,\widetilde{\sigma}}\widetilde{E}_{k,j,\widetilde{\sigma}}c^{\dagger}_{k,j,\widetilde{\sigma}}c_{k,j,\widetilde{\sigma}}+
\sum_{q}E^{ph}_{q}b^{\dagger}_{q}b_{q} +
\sum_{k,q}[g_{++}(k,q)\times \notag \\ &\times c^{\dagger}_{k,1,+}c_{k+q,2,+} +
g_{--}(k,q)c^{\dagger}_{k,1,-}c_{k-q,2,-} +
g_{+-}(k,q)\times \label{Hfinal} \\ \notag &\times c^{\dagger}_{k,1,+}c_{k+q,2,-} + g_{-+}(k,q)c^{\dagger}_{k,1,-}c_{k+q,2,+}]b^{\dagger}_{q} + H.c.,
\end{align}
where $\widetilde{E_{j}}(k)$ stands for the standard Rashba dispersions
\begin{equation}
\widetilde{E_{j}}(k)=E_{j}(k)\pm\alpha^{R}_{j}|k|=\frac{\hbar^{2}k^{2}}{2m}\pm\alpha^{R}_{j}|k|
\label{ERashba}
\end{equation}
shown schematically in Fig. \ref{Fig1}(b) for upper and lower subband $j$.

The electron-photon coupling coefficients now are functions of electron momentum $k$ and are spin-dependent:
\begin{align}
g_{++}(k,q)=g_{q}e^{-i\frac{\theta_{k}-\theta_{k+q}}{2}}\cos\left(\frac{\theta_{k}-\theta_{k+q}}{2}\right)=g_{q}\widetilde{g}_{++}, \label{gpp}\\
g_{+-}(k,q)=ig_{q}e^{-i\frac{\theta_{k}-\theta_{k+q}}{2}}\sin\left(\frac{\theta_{k}-\theta_{k+q}}{2}\right)=g_{q}\widetilde{g}_{+-}, \\
g_{-+}(k,q)=ig_{q}e^{-i\frac{\theta_{k}-\theta_{k+q}}{2}}\sin\left(\frac{\theta_{k}-\theta_{k+q}}{2}\right)=g_{q}\widetilde{g}_{-+}, \\
g_{--}(k,q)=g_{q}e^{-i\frac{\theta_{k}-\theta_{k+q}}{2}}\cos\left(\frac{\theta_{k}-\theta_{k+q}}{2}\right)=g_{q}\widetilde{g}_{--},\label{gmm}
\end{align}
where $g_{q}$ is an absolute value of light-matter coupling coefficient defined above.

Note, that for non-zero photon momentum $q\neq0$ the spin-flip transitions become possible ($\widetilde{g}_{+-},~\widetilde{g}_{-+}\neq0$) and all four transitions with different energies are allowed (see Fig. \ref{Fig1}(b)). However, the main impact for photoabsorption comes from electrons situated close to the Fermi surface, for which $q\ll k_{F}$. Hence spin-flip matrix elements are proportional to the momenta ratio $\widetilde{g}_{+-},~\widetilde{g}_{-+}\sim q/k_F \approx 0$, their oscillator strength is small and only two transitions $+ \rightarrow +$ and $- \rightarrow -$ can be expected to be visible in the experiment. Note, that the energies of these transitions are different, and thus they can be excited selectively by tuning the frequency of the excitation beam.

\subsubsection{Rashba and Dresselhaus SOI}
In the realistic QWs Rashba SOI is not the only interaction acting on the electron spin since due to bulk asymmetry the Dresselhaus SOI term is usually present in the Hamiltonian. The combination of both Rashba and Dresselhaus SOI leads to spin anisotropy and peculiar spin orientation in the subbands. The total electronic Hamiltonian $H=H_{SIA}+H_{BIA}$ (see Eqs. (\ref{HRashba}) and (\ref{HDresselhaus})) can be diagonalized in the same fashion as above by introduction of the operators
\begin{eqnarray}
c_{k,\pm}=\frac{1}{\sqrt{2}}\left[\pm\frac{i\alpha e^{-i\theta_{k}}+\beta e^{i \theta_{k}}}{\sqrt{\alpha^{2}+\beta^{2}+2\alpha\beta\sin\theta_{k}}}a_{k,\uparrow}+a_{k,\downarrow}\right].
\label{new_operators_DR}
\end{eqnarray}
The dispersion relations for electrons are not cylindrically symmetric and have non-trivial form
\begin{equation}
\widetilde{E_{j}}(k)=E_{j}(k)\pm|k|\sqrt{\alpha^{2}_{j}+\beta^{2}_{j}+2\alpha\beta\sin 2\theta_{k}},
\label{EDress}
\end{equation}
where $j$ denotes the number of subband.

In full analogy with the case when only Rashba SOI is present, one can get the matrix elements for all four optical transitions. While the calculation is straightforward, their explicit expressions are rather cumbersome and we do not present them here. Similarly to the case of isotropic Rashba SOI the spin-flip transitions are suppressed, $\widetilde{g}_{-+},~\widetilde{g}_{+-}\approx 0$.

\subsection{Spin polarization}
\label{subsec:polarization}

\subsubsection{Rashba SOI}

The achievement of non-zero spin polarization is one of the main goals of spintronics. Regarding the optical generation of spin polarization (spin orientation), there are many proposals based on the interband excitation, in particular by circularly polarized light.\cite{Meier} As well, there are several proposals for spin generation with intersubband transition, where optical selection rules imply that only linearly polarized light is absorbed and achieving of non-zero spin polarization becomes a formidable task. One mechanism of generation is based on different strength of $+ \rightarrow +$ and $- \rightarrow -$ transitions due to valence band mixing which modifies transition matrix elements.\cite{TarasenkoJETP} Another approach requires different SOI effective field in both fundamental and excited subband leading to overall non-zero spin polarization.\cite{TarasenkoPSS} Here we want to study the possibility of spin polarization generation which is linearly proportional to the photon wave vector $q$.

The average spin of the electron in the second quantization representation reads \cite{Bruus,Mahan}
\begin{equation}
\langle\hat{S}\rangle=\sum_{k\sigma \sigma'}\langle k\sigma'|\hat{S}|k\sigma \rangle c^{\dagger}_{k,\sigma '}c_{k,\sigma},
\label{spinoperator}
\end{equation}
where we defined the spin operator $\hat{S}=\frac{\hbar}{2}(\sigma_{x},\sigma_{y},\sigma_{z})$ with $\sigma_{i}$ being Pauli matrices given for each direction ($i=x,y,z$). It is reminiscent to the first quantization definition of average spin given as $S_{k,\pm}=\langle\Psi_{k,\pm}|\hat{\sigma}|\Psi_{k,\pm}\rangle$.\cite{AverkievGlazov}

\begin{figure}
\includegraphics[width=1.0\linewidth]{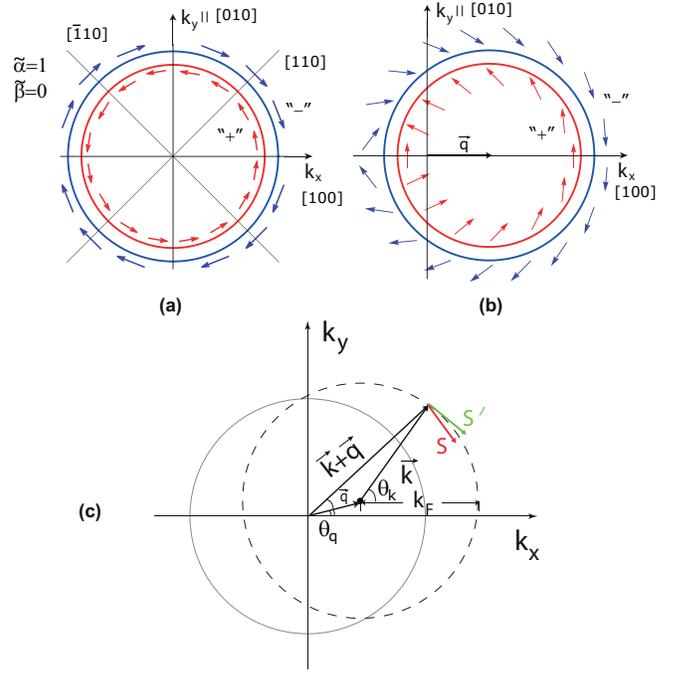}
\caption{(Color online) (a) Orientation of electron spin in $+$ and $-$ Rashba splitted subbands without accounting of photon momentum ($q=0$). (b) Vector plot of spin orientation with accounting of finite photon momentum ($q/k_{F}=0.3$, $\theta_{q}=0$). (c) Schematic representation of non-zero photon momentum $q$ influence on spin polarization.}
\label{Fig2}
\end{figure}

First, let us consider the case of fully thermalized electron gas and calculate the average spin within subband. The average spin projection onto $x$ and $y$ directions for the case of Rashba SOI are given by
\begin{align*}
&\langle +|S_{x}|+ \rangle=\sum_{\mathbf{k}}\sin\theta_{k}=\int_{0}^{k_{+}}\frac{kdk}{(2 \pi)^{2}}\int_{0}^{2 \pi}d\theta_{k}\sin\theta_{k},\\
&\langle +|S_{y}|+ \rangle=\sum_{\mathbf{k}}(-\cos\theta_{k})=\int_{0}^{k_{+}}\frac{kdk}{(2 \pi)^{2}}\int_{0}^{2 \pi}d\theta_{k}(-\cos\theta_{k}),\\
&\langle -|S_{x}|- \rangle=\sum_{\mathbf{k}}(-\sin\theta_{k})=\int_{0}^{k_{-}}\frac{kdk}{(2 \pi)^{2}}\int_{0}^{2 \pi}d\theta_{k}(-\sin\theta_{k}),\\
&\langle -|S_{y}|- \rangle=\sum_{\mathbf{k}}\cos\theta_{k}=\int_{0}^{k_{-}}\frac{kdk}{(2 \pi)^{2}}\int_{0}^{2 \pi}d\theta_{k}\cos\theta_{k},
\end{align*}
and one can see the orientation for spin in $++$ and $--$ transitions given by vectors $\mathbf{S}_{+}=(\sin\theta_{k},-\cos\theta_{k},0)$ and $\mathbf{S}_{-}=(-\sin\theta_{k},\cos\theta_{k},0)$ with three
spatial components $\mathbf{S}=(S_x,S_y,S_z)$. The modified radii of Fermi surface for $+$ and $-$ spin subbands are written as $k_{\pm}=k_{F}(1 \mp \eta_{R})$ (see Fig. \ref{Fig1}(b)), where $\eta_{R}=\frac{\alpha m}{\hbar^{2}k_{F}}$. The electron spin orientation for spin subbands is shown in Fig. \ref{Fig2}(a). Obviously, the integration over angle $\theta_{k}$ yields zero average spin projections, $\langle \pm |S_{x,y}|\pm \rangle = 0$.

Now, let us consider the situation when electrons are constantly optically excited to the upper subband due to cavity photon
absorption. The spin of an electron in the presence of spin-orbital interaction of Rashba type is perpendicular to its momentum direction. Therefore, with accounting of finite wave vector of the photon one can see that spin state in Fig. \ref{Fig2}(c) is no longer given by vector $\mathbf{S}^{'}$, but $\mathbf{S}$. This corresponds to the change of spin states to $\mathbf{S}_{+}=(\sin\theta_{k+q},-\cos\theta_{k+q},0)$ and $\mathbf{S}_{-}=(-\sin\theta_{k+q},\cos\theta_{k+q},0)$. Now the boundaries of integration are changed due to shift of Fermi circles and for selective excitation of $++$ and $--$ transitions are given by expression
\begin{equation}
k'_{\mp}=k_{\mp}\Big[\sqrt{1-\frac{q^2}{k_{\mp}^{2}}\sin^{2}(\theta_{k}-\theta_{q})}+\frac{q}{k_{\mp}\cos(\theta_{k}-\theta_{q})}\Big].
\label{kprimeR}
\end{equation}

The average spin projections with accounting of photon momentum can be rewritten using relations
\begin{align}\notag
&\cos\theta_{k+q}=\frac{k\cos\theta_{k}+q\cos\theta_{q}}{\sqrt{k^{2}+q^{2}+2kq\cos(\theta_{k}-\theta_{q})}},\\
&\sin\theta_{k+q}=\frac{k\sin\theta_{k}+q\sin\theta_{q}}{\sqrt{k^{2}+q^{2}+2kq\cos(\theta_{k}-\theta_{q})}},
\label{geometry}
\end{align}
where $\theta_{q}$ denotes angle between vector $\mathbf{q}$ and $x$ axis, and all variables are shown explicitly on the sketch in Fig. \ref{Fig2}(c). One should keep in mind that Fermi circle for $q \neq 0$ is $\theta_{k}$-dependent. Therefore, it requires advanced momentum integration with consequent angular integration and can be done only numerically. To obtain the dimensionless quantity connected to average spin of electron gas (or spin polarization) one should normalize the result of integration dividing it by population in each subband $n_{\pm}=\frac{k_{\pm}^{2}}{4\pi}$.

One sees that while for $q=0$ case average spin is zero, the accounting of Fermi circle shift gives preferable direction to the spin of excited electrons which is $[010]$ for $\theta_{q}=0$ incident angle shown in Fig. \ref{Fig2}(b). This effect is similar to spin-galvanic effect observed previously for interband case and connected to non-zero charge current.\cite{Silsbee,GanichevNature}

\subsubsection{Rashba and Dresselhaus SOI}

Following the scheme used in the previous subsection for the spin-isotropic Rashba splitting, we derive the same quantities for spin-anisotropic case of combined Rashba and Dresselhaus interaction.

\begin{figure}
\includegraphics[width=1.0\linewidth]{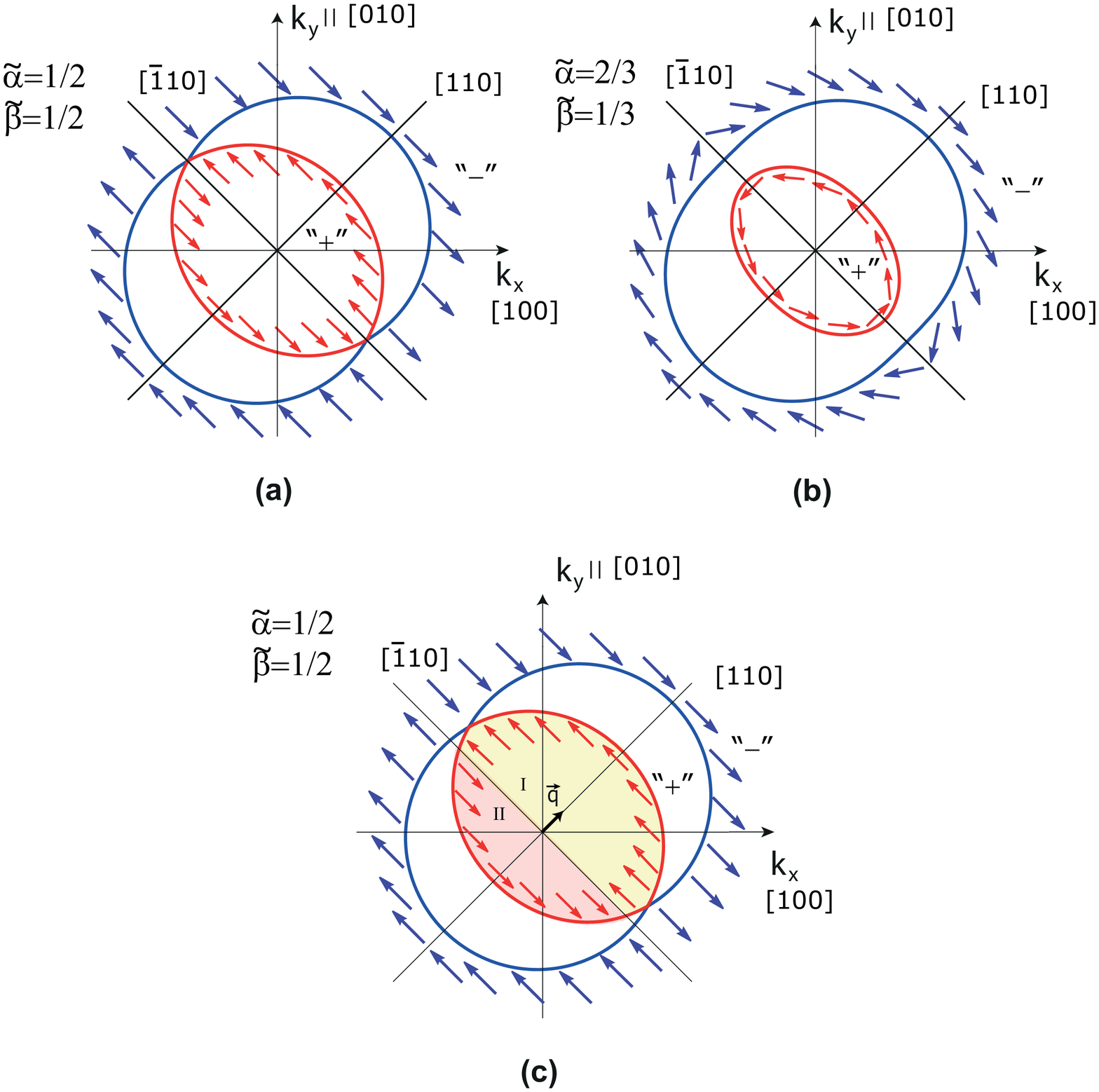}
\caption{(Color online) Orientation of electron spin in $+$ and $-$ Rashba and Dresselhaus splitted subbands for equal SOI strength (a) and for $\widetilde{\alpha}=2/3$, $\widetilde{\beta}=1/3$ case (b). (c) The spin vector plot for equal Rashba and Dresselhaus SOI with accounting of finite photon momentum which leads to Fermi surface shift in $[110]$ direction. The areas $I$ (yellow filling) and $II$ (pink filling) show the difference in spin orientation.}
\label{Fig3}
\end{figure}

The spin orientation can be found using the spin states for Rashba and Dresselhaus case (\ref{new_operators_DR}) and is given by
\begin{align*}
&\langle +|S_{x}|+ \rangle=\sum_{\mathbf{k}}\frac{\widetilde{\alpha}\sin\theta_{k}+\widetilde{\beta}\cos\theta_{k}}{\sqrt{\widetilde{\alpha}^{2}+\widetilde{\beta}^{2}+2\widetilde{\alpha}\widetilde{\beta}\sin 2\theta_{k}}},\\
&\langle +|S_{y}|+ \rangle=\sum_{\mathbf{k}}\frac{-\widetilde{\alpha}\cos\theta_{k}-\widetilde{\beta}\sin\theta_{k}}{\sqrt{\widetilde{\alpha}^{2}+\widetilde{\beta}^{2}+2\widetilde{\alpha}\widetilde{\beta}\sin 2\theta_{k}}},\\
&\langle -|S_{x}|- \rangle=\sum_{\mathbf{k}}\frac{-\widetilde{\alpha}\sin\theta_{k}-\widetilde{\beta}\cos\theta_{k}}{\sqrt{\widetilde{\alpha}^{2}+\widetilde{\beta}^{2}+2\widetilde{\alpha}\widetilde{\beta}\sin 2\theta_{k}}},\\
&\langle -|S_{y}|- \rangle=\sum_{\mathbf{k}}\frac{\widetilde{\alpha}\cos\theta_{k}+\widetilde{\beta}\sin\theta_{k}}{\sqrt{\widetilde{\alpha}^{2}+\widetilde{\beta}^{2}+2\widetilde{\alpha}\widetilde{\beta}\sin 2\theta_{k}}},
\end{align*}
where we defined dimensionless Rashba and Dresselhaus constants $\widetilde{\alpha}=\frac{\alpha}{\alpha+\beta}$ and $\widetilde{\beta}=\frac{\beta}{\alpha+\beta}$, respectively. The vector plot of spin orientation for different Rashba and Dresselhaus constants is shown in Fig. \ref{Fig3} (a, b). One sees that while for $\widetilde{\alpha}=2/3$, $\widetilde{\beta}=1/3$ the spin pattern is close to Rashba case but with preferred directions being $[\overline{1}10]$ and $[\overline{1}\overline{1}0]$, for equal Rashba and Dresselhaus SOI strength ($\widetilde{\alpha}=\widetilde{\beta}=1/2$) the spin orientation approaches step-like function which changes sign in $\theta_{k}=3 \pi/4$ and $\theta_{k}=7 \pi/4$ points. The Fermi surfaces for upper ($+$) and lower ($-$) spin subbands now are given by expression $k_{\pm}=k_{F}(1\mp \eta(\theta_{k}))$ with $\eta(\theta_{k})=\frac{m}{\hbar^{2}k_{F}}\sqrt{\alpha^{2}+\beta^{2}+2\alpha\beta\sin 2\theta_{k}}$.

\begin{figure}
\includegraphics[width=0.85\linewidth]{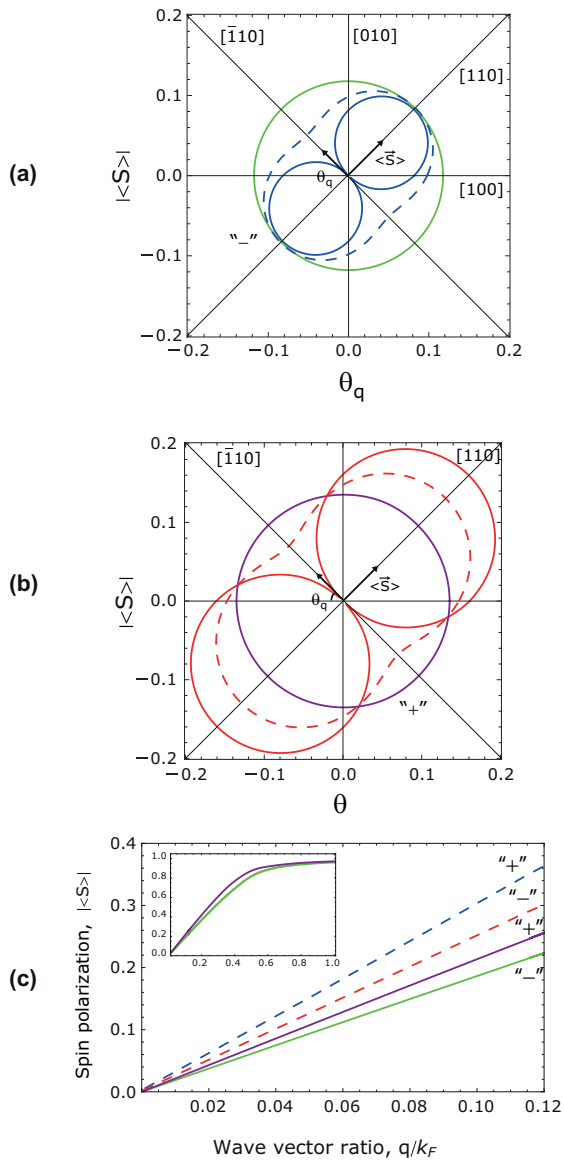}
\caption{(Color online) (a) The polar plot of absolute value of an average spin vector as a function of photon incident angle $\theta_{q}$ for $++$ transition. Green solid circle corresponds to the Rashba SOI only case ($\widetilde{\alpha}=1$,
$\widetilde{\beta}=0$). Blue solid circles show strong anisotropic dependence of spin polarization on incident angle $\theta_{q}$ for equal Rashba and Dresselhaus strength ($\widetilde{\alpha}=\widetilde{\beta}=1/2$). The blue dashed line
shows transient behaviour for $\widetilde{\alpha}=2/3$ and $\widetilde{\beta}=1/3$ case. (b) The polar plot of absolute value of average spin vector as a function of photon incident angle $\theta_{q}$ for the lower $--$ spin transition. Purple solid circle corresponds to the isotopic Rashba SOI case ($\widetilde{\alpha}=1$, $\widetilde{\beta}=0$). Red solid circles show spin polarization as a function of incident angle $\theta_{q}$ for equal Rashba and Dresselhaus strength ($\widetilde{\alpha}=\widetilde{\beta}=1/2$). The red dashed line shows transient behaviour for $\widetilde{\alpha}=2/3$ and $\widetilde{\beta}=1/3$ case. All dependences are calculated for $q/k_{F}=0.06$. (c) The spin polarization dependence on the wave vector ratio $q/k_{F}$ for the Rashba case (green and purple solid lines) and Rashba with Dresselhaus case (blue and red dashed lines) for $\widetilde{\alpha}=2/3$ and $\widetilde{\beta}=1/3$. The inset shows the behaviour of spin polarization for Rashba only case on large scale of argument.} \label{Fig4}
\end{figure}

Accounting for non-zero photon momentum result into change of spin states $\mathbf{S}(\theta_{\mathbf{k}})\rightarrow \mathbf{S}^{'}(\theta_{\mathbf{k}+\mathbf{q}})$. Additionally, one should account for angle dependent shifted Fermi surfaces
\begin{align}\notag
&k'_{\pm}=k_{F}(1+\eta(\theta_{k,q}))\Big[\sqrt{1-\frac{q^2}{k_{F}^{2}(1+\eta(\theta_{k,q}))^{2}}\sin^{2}(\theta_{k}-\theta_{q})}+\\
&+\frac{q}{k_{F}(1+\eta(\theta_{k,q}))}\cos(\theta_{k}-\theta_{q})\Big],
\label{kprimeDR}
\end{align}
with $\eta(\Theta_{k,q})=\frac{m}{\hbar^{2}k_{F}}\sqrt{\alpha^{2}+\beta^{2}+2\alpha\beta\sin 2\Theta_{k,q}}$ and $\Theta_{k,q}=\arctan \Big[ \frac{k\sin\theta_{k}-q\sin\theta_{q}}{k\cos\theta_{k}-q\cos\theta_{q}} \Big]$. Similarly to the isotropic Rashba case, for $q=0$ the average spin vanishes, while for finite photon wave vector $q$ the net spin is not equal to zero. The effect is most clear for equal Rashba and Dresselhaus SOI. Then the average spin can be seen as difference of areas $I$ and $II$ in Fig. \ref{Fig3}(c). Moreover, one can note that $\widetilde{\alpha}=\widetilde{\beta}=1/2$ case is strongly anisotropic regarding the photon incidence direction $\theta_{q}$ (Fig. \ref{Fig4}(a, b)). While for $\theta_{q}=\pi/4$ it has maximum absolute value, spin polarization does not appear in the system for $\theta_{q}=3 \pi/4$ angle of incidence. The average spin vector $\langle \vec{S} \rangle$ is oriented perpendicular to $\mathbf{q}$. Dashed lines in Fig. \ref{Fig4}(a, b) correspond to not equal Rashba and Dresselhaus SOI and depict transient regime between spin-anisotropic to spin-isotropic case.

We calculate the dependence of spin polarization on the orientation of an incident light angle $\theta_{q}$. The corresponding polar plots are represented in Fig. \ref{Fig4} for selective excitation of $--$ (a, green solid circle) and $++$ (b, purple solid circle) transitions. Note, that in case where only Rashba term is present, the absolute value of average spin is the same for all $\theta_{q}$. The orientation of average spin vector is defined by the $\theta_{q}$ angle and is aligned perpendicular to the vector $\mathbf{q}$ (Fig. \ref{Fig4}, $\langle\vec{S}\rangle$ label on the plot).

Furthermore, we study the dependence of spin polarization on the wave vector ratio $q/k_{F}$ which defines the relative value of the Fermi circle shift under optical excitation. Since photon wave vector value usually does not overcome inverse micrometer $q<10^{-6}$ m$^{-1}$, the experimental way to tune $q/k_{F}$ ratio is changing the carrier density. The plot of net spin polarization as a function of momenta ratio $q/k_{F}$ is given in Fig. \ref{Fig4}(c) for $--$ and $++$ transitions (green and purple solid lines). Higher values of spin polarization for lower spin subband come from the fact that $k_{-}<k_{F}<k_{+}$. One can see the linear dependence for the case of isotropic SOI for the realistic values of
wave vector ratio. In the inset of Fig. \ref{Fig4}(c) we show the behavior of average spin on the large scale. It shows saturation for region where photon wave vector approaches the value of Fermi circle radius with spin polarization approaching unity. However, firstly, this situation does not have substantial physical background since for small concentrations ($q\approx k_{F} \approx 10^{-6}$ m$^{-1}$) even small excitation leads to the strongly non-equilibrium situation and this situation can not be treated within our approach. Secondly, the visible separation of polaritonic
modes implies that separation between bare transition modes $\epsilon_{s}=2(\alpha_{2}-\alpha_{1})k_{F}$ is greater than photon decay rate ($\Gamma \approx 1$ meV), which puts the lower boundary for $k_{F}$ value. Finally, for the ratio $q/k_{F}$ closer to unity one should account for $+-$ and $-+$ transitions on the same footing with $++$ and $--$ transitions, and we do not consider this situation.

\subsection{Spin currents}
\label{subsec:current}

In the following section we briefly discuss the possibility of spin current generation by intersubband optical excitation. The concept of spin currents lies in the origin of spintronics since it allows to design devices based on manipulation of spin degree of freedom of the carriers. However, effective creation of spin currents is not a trivial task. The proposed schemes include injection from ferromagnetic leads to semiconductor,\cite{Zutic} creation of non-equilibrium distribution of carriers by applying voltage\cite{Silsbee} or by optical excitation \cite{TarasenkoSST,TarasenkoJETPLett,Sherman} in structures with spin-orbit interaction. The photocurrents associated with non-zero spin polarization were studied in the context of interband transition with spin-split bands shined by circularly polarized light.\cite{Lyanda-Geller89,Lyanda-Geller90,Bhat} The inverse effect namely the appearance of charge current due to non-equilibrium population of electron spins was observed by Ganichev \textit{et al.}\cite{GanichevNature} and named spin-galvanic effect. The review on the subject can be found in Ref. [\onlinecite{GanichevReview}].

\begin{figure}
\includegraphics[width=1.0\linewidth]{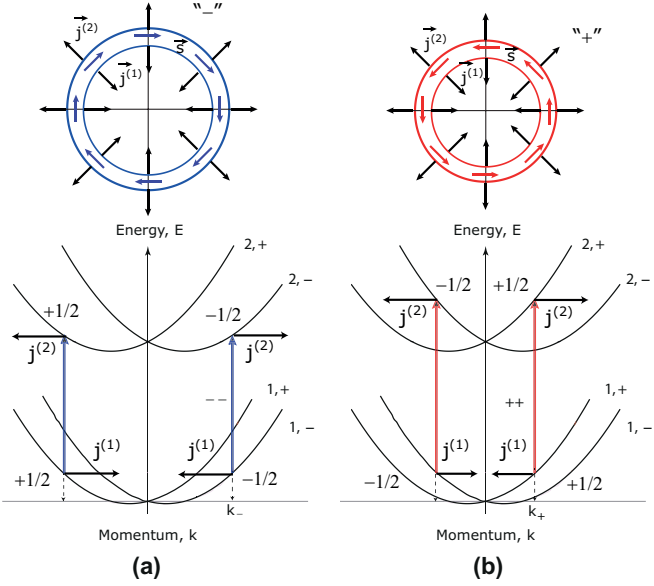}
\caption{(Color online) Sketch of spin current orientation ($\vec{j}$, black arrows) for the $--$ (a) and $++$ (b) transition. The coloured lines show the spin orientation of electrons ($\vec{s}$).}
\label{Fig5}
\end{figure}

In the system with spin-split upper and lower subbands the intersubband transitions occur with spin conservation. Due to the
difference in the SOI constants for upper and lower subband, the $++$ and $--$ transitions have different energies. An absorption of TM polarized photon leads to the electron in the upper subband with momentum $k$ (first we consider vertical transition) and ``hole'' in the lower subband with the opposite momentum (see Fig. \ref{Fig5}). Moreover, we account for the Rashba SOI with peculiar spin orientation in the spin subbands. The excitation of carries on the Fermi disc leads to the generation of spin and charge currents (Fig. \ref{Fig5}, top). However, for the case of vertical transition charge current is zero and situation corresponds to the generation of \emph{pure} spin currents.  The expression for the pure spin current in the general form can be described by the pseudotensor $\widehat{\mathbf{J}}$ with components\cite{TarasenkoSST}
\begin{equation}
J_{\mu}^{\nu}=\sum_{k}\tau_{e}Tr\Big[\frac{\sigma_{\mu}}{2}\mathbf{v}_{\nu}\dot{\rho}(\mathbf{k})\Big],
\label{spin_current}
\end{equation}
where $\mathbf{v}(\mathbf{k})$ is velocity operator and $\rho(\mathbf{k})$ is spin density matrix, $\tau_{e}$ is relaxation
rate. This formula applied to the intersubband transition describes the spin currents in $\nu$ direction with spin aligned in $\mu$ direction.

In the previous section we have shown that accounting for the finite photon momentum leads to the non-zero average spin of intersubband excitation. Thus, it creates a non-equilibrium situation similar to the electric field displacement of Fermi disks,\cite{Silsbee} when the charge current in the system does not vanish and spin current with spin polarization defined by incident angle $\theta_{q}$ is generated. While we do not derive the expressions for spin current in $q\neq 0$ case, the behavior can be qualitatively understood in connection with previous section.

Finally, the implementation of strong-coupling between photonic mode and intersubband excitation affects the spin currents. Due to coherence of cavity mode the resulted spin currents will have the long coherence length and can be used for spinoptronic applications.\cite{Glazov}

\section{Intersubband polaritons with SOI}
\label{sec:polaritons}

\subsection{Elementary excitation spectrum}

\begin{figure}
\includegraphics[width=1.0\linewidth]{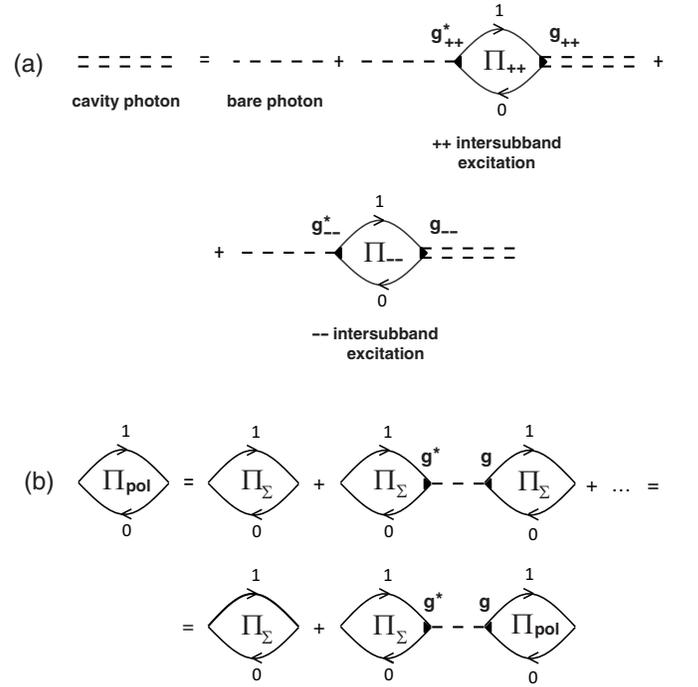}
\caption{Diagrammatic representation of intersubband polaritons. (a), The Dyson equation for Green function of microcavity photon interacting with $++$ and $--$ intersubband transitions $\Pi_{\pm,\pm}$. $g_{\pm}$ corresponds to the interaction constant, indices $1$ and $0$ denote the excited and fundamental subband, respectively. (b) Dyson equation for polarization operators governing the absorbtion spectrum of intersubband quasiparticles coupled to the cavity mode where $\Pi_{\Sigma}=\Pi_{++}+\Pi_{--}$.} \label{Fig6}
\end{figure}

In the case when QW is placed inside a microcavity which is tuned in resonance with intersubband transition, the photons can undergo multiple re--emissions and re--absorptions and hybrid intersubband polariton modes can be formed.\cite{Dini,Geiser} The powerful theoretical tool for their description is many body diagrammatic technique.\cite{Kyriienko} The quantities which define the physical properties of the system are photon Green function and polarization operator, shown diagrammatically in Fig. \ref{Fig6}(a) and Fig. \ref{Fig6}(b), respectively. The poles of the renormalized photon Green function give the spectrum of elementary excitations in the system, while imaginary part of polarization operator allows to calculate the absorption spectrum.

If one neglects the electron-electron interactions, the polarization operators corresponding for 4 possible types of transitions in the system ($++,--,+-,-+$) read
\begin{equation}
\Pi_{0}^{\widetilde{\sigma}\widetilde{\sigma}^{'}}(\omega,q)=\int\frac{d\mathbf{k}}{(2\pi)^{2}}\frac{|\widetilde{g}^{2}_{\widetilde{\sigma}\widetilde{\sigma}^{'}}(q,\varphi)|}{\hbar\omega+E_{\mathbf{k},\widetilde{\sigma}}^{(1)}-E_{\mathbf{k+q},\widetilde{\sigma}^{'}}^{(2)}+i\gamma},
\label{P01}
\end{equation}
where we defined $\varphi=\theta_{k}-\theta_{k+q}$ and $\widetilde{\sigma},\widetilde{\sigma}^{'}=\pm$ denote different
spin subbands ($+$ and $-$). The energy of electron in fundamental subband is $E_{\widetilde{\sigma}}^{(1)}(k)=\hbar^{2}k^{2}/2m\pm \alpha_{1}|k|$ and in upper subband is $E_{\widetilde{\sigma}}^{(2)}(k+q)=\Delta+\hbar^{2}(k+q)^{2}/2m\pm \alpha_{2}|k+q|$ and dimensionless coefficients $\widetilde{g}_{\widetilde{\sigma}\widetilde{\sigma}^{'}}(q,\varphi)$ are defined by Eqs. (\ref{gpp})--(\ref{gmm}) and read
\begin{align}\label{gcoeffs}
&|\widetilde{g}^{2}_{++}(q,\varphi)|=|\widetilde{g}^{2}_{--}(q,\varphi)|=\cos^{2}\varphi/2=\frac{1}{2}(1+\cos\varphi), \\
&|\widetilde{g}^{2}_{+-}(q,\varphi)|=|\widetilde{g}^{2}_{-+}(q,\varphi)|=\sin^{2}\varphi/2=\frac{1}{2}(1-\cos\varphi).
\notag
\end{align}
It was already mentioned in Sec. \ref{subsec:model} that while $|\widetilde{g}^{2}_{\pm\pm}|$ interaction constants corresponding to $++,~--$ transitions are close to unity, the $+-,~-+$ transitions do not play substantial role. The quantitative estimate can be done using realistic values of electron concentration with Fermi momentum of the order $k_{F}\approx 10^8$ m$^{-1}$ ($n=10^{11}$ cm$^{-2}$) and microcavity photon momentum $q$ in the point of anticrossing
$q=10^6$ m$^{-1}$ for negative detuning $\delta=-5$ meV. The angle is typically small $\varphi=\theta_{k}-\theta_{k+q}\approx \arcsin(q/k_{F})\approx 10^{-2}$ which reduces the oscillator strength of spin-flip transitions $\widetilde{g}^{2}_{\widetilde{\sigma}\widetilde{\sigma}^{'}}(q,\varphi)$ by five orders of magnitude. Consequently, they give minor contribution in the case of strong light-matter coupling, and we omit them in the Dyson equation for cavity photon Green function shown in Fig. \ref{Fig6}(a). The polarization operators corresponding to $++$ and $--$ transitions can be calculated by performing analytical integration on angle in (\ref{P01}) with subsequent numerical integration on absolute value of wave vector $k$.

The dispersion of elementary particles in the system is defined by the poles of renormalized photon Green function $D(q,\omega)$, which can be found by solving the Dyson equation, and reads
\begin{equation}
D=\frac{D_{0}}{1-g_{q}^2(\Pi_{++}+\Pi_{--})D_{0}}, \label{D}
\end{equation}
where $\Pi_{\pm,\pm}$ are polarization operators described before, $q_q$ is given by Eq. (\ref{g}) and $D_0$ is a bare photon Green function,
\begin{equation}
D_{0}(\omega,q)=\frac{2\hbar\omega_{0}(q)}{\hbar^{2}\omega^{2}-\hbar^{2}\omega_{0}^{2}(q)+2i\Gamma\omega_{0}(q)},
\label{D0}
\end{equation}
where $\omega_{0}(q)$ denotes cavity mode dispersion and $\Gamma$ is a broadening of photonic mode due to the finite lifetime of the cavity photon (taken to be $\approx 10$ ps). The explicit expression for the electron-photon interaction matrix element $g_{q}$ is given by Eq. (\ref{g}).

\begin{figure}
\includegraphics[width=1.0\linewidth]{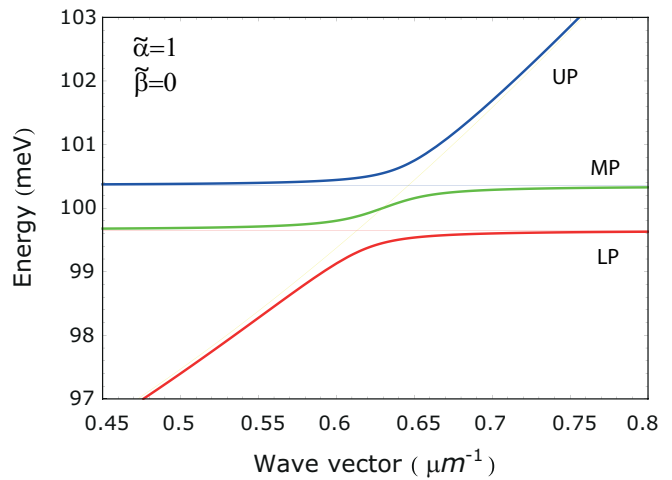}
\caption{(Color online) Dispersions of intersubband polariton modes with accounting of isotropic Rashba splitting. The red line corresponds to lower polariton (LP) and coincides with $--$ excitation dispersion for large momenta, while blue line shows upper polariton (UP) which is reminiscent to $++$ excitation for small wave vectors. The green line describes middle or mixed polariton (MP) being a transitional branch between lower and upper polaritonic branch.}
\label{Fig7}
\end{figure}

The spectrum of the elementary excitations of intersubband transitions coupled to the cavity mode with accounting of Rashba SOI is plotted in Fig. \ref{Fig7}(a). Here we considered AQW of $L=12.8$ nm width with frequency of bare transition equal to $\Delta=100$ meV at $k=0$ point and concentration of electron gas $n=10^{11}$ cm$^{-2}$. The Rashba constants for lower and upper subband were taken to be $\alpha_{1}=0.9$ meV nm and $\alpha_{2}=6$ meV nm, respectively,\cite{Sherman} and we neglected Dresselhaus terms for simplicity. Both $++$ and $--$ spin-dependent transitions are in strong-coupling regime with the cavity mode. As transition frequencies are different, one observes the formation of three polariton branches. This differs from the usually considered spin-independent case for which only two polaritonic branches exist.

The energy splitting between $++$ and $--$ transitions is approximately $\epsilon_{s}=2(\alpha_{2}-\alpha_{1})k_{F}$. For
instance, with the given parameters $\alpha_{2}$ and $\alpha_{1}$ and for concentration $n=10^{11}$ cm$^{-2}$ it has value
$\epsilon_{s}=0.8$ meV, which is already comparable to the broadening of the photonic mode. Therefore, for the experimental observation of the effect one should search for the sample with large difference in Rashba constants, and InAs based QWs are promising candidates for this purpose.

While the consideration of simultaneous Rashba and Dresselhaus interaction can be done in the same fashion, it does not lead to qualitative differences comparing to the isotropic Rashba case and we do not address this point here.

\begin{figure}
\includegraphics[width=1.0\linewidth]{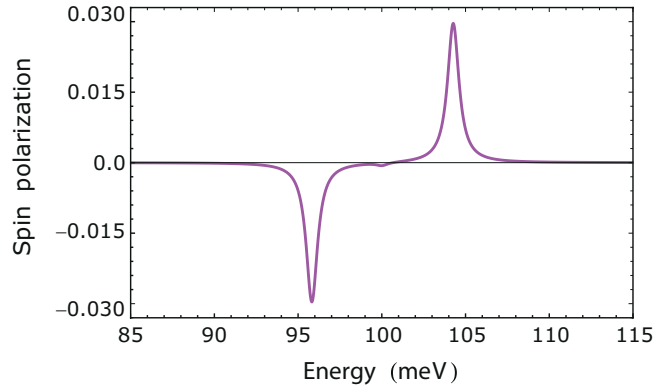}
\caption{(Color online) The spin polarization degree as a function of pumping laser energy for the photon wave vector $q=6.25$ $\mu$m$^{-1}$. Changing the laser frequency allows one to excite selectively upper, middle or lower polariton modes. This results into change of the net spin polarization.}
\label{Fig8}
\end{figure}

Intersubband polariton modes can be selectively excited by the resonant excitation of a given frequency. As it was discussed above, such excitation will induce spin currents in the system. The net spin polarization as a function of the excitation frequency is shown in Fig. \ref{Fig8}.

Due to the coupling with photonic mode, spin currents induced in the coupled QW--microcavity system will have much bigger decoherence length compare to the electron currents.\cite{Langbein} This makes the considered system a promising candidate for spintronic and spinoptronic applications.

\section{Conclusions}
\label{sec:conclusions} In conclusion, we analyzed the optical properties of spin-dependent intersubband transition in asymmetric quantum wells. We have shown that accounting of finite photon momentum leads to the optical orientation effect. We also calculated the spectrum of elementary excitations arising from strong coupling of the photonic mode with an intersubband transition of an asymmetric QW. The calculated spectrum of elementary excitations show the appearance of upper, lower, and middle polaritonic modes. The possibility of the generation of spin currents with long coherence lengths is discussed.

We would like to thank M. M. Glazov and S. A. Tarasenko for useful discussions and the critical reading of the manuscript. This work was supported by Rannis ``Center of Excellence in Polaritonics'' and FP7 IRSES project ``SPINMET''. O.K. acknowledges the support from Eimskip Fund.


\end{document}